\documentclass{article}
\usepackage{amsmath}
\usepackage{amssymb}
\usepackage{graphicx}
\begin{document}

\begin{center}
\large
\textbf{Local tomography and the role of the complex numbers}
\\[0,1 cm]
\textbf{in quantum mechanics}
\\[2,0 cm]
\normalsize
Gerd Niestegge
\\[0,3 cm]
\footnotesize
gerd.niestegge@web.de
\\[2,0 cm]
\end{center}
\normalsize

\begin{abstract}
Various reconstructions of finite-dimensional quantum mechanics
result in a formally real Jordan algebra $A$ and a last step remains 
to conclude that $A$ is the self-adjoint part of a C*-algebra.
Using a quantum logical setting, it is shown that this can be achieved 
by postulating that there is a 
locally tomographic model for a composite system consisting of two copies
of the same system. 
Local tomography is a feature of classical probability theory
and quantum mechanics; it means that state tomography
for a multipartite system can be performed by 
simultaneous measurements in all subsystems.
The quantum logical definition of local tomography 
is sufficient, but it is less restrictive than the prevalent definition 
in the literature and involves some subtleties
concerning the so-called spin factors.
\\[0,5 cm]
\\[0,5 cm]
\\[1,5 cm]
\end{abstract}

\noindent
\large
\textbf{1. Introduction}
\normalsize
\\[0,3 cm]
The quantum-mechanical need for
the mathematical Hilbert space apparatus
including the complex numbers 
has been a matter of fundamental research
since the emergence of this theory 
one hundred years ago. 
Over time, quantum logical, algebraic, operational and information theoretic approaches
to reconstruct quantum mechanics 
from a small number of plausible principles
have been proposed, using algebraic methods,
generalized probabilistic theories or category theory.
Particular emphasis has been placed 
on information theoretic principles
in the last two decades.

Several approaches 
\cite{Barnum_2013, barnum2019strongly, e19060253, 
barnum2014higher, nakahira2019derivation, 
nie2020charJordan, wetering2018, van_de_Wetering_2019, 
Wilce_2012, Wilce_2018, Wilce_2019}
succeed in deriving the need for
formally real Jordan algebras.
A further approach \cite{nie2004why, nie2012AMP}
also results in the formally real Jordan algebras,
when it is restricted to the finite-dimensional case.
This type of Jordan algebra includes
quantum mechanics with the complex numbers, 
but also other versions with the real numbers,
quaternions, octonions and mixtures of these versions.

One feature that distinguishes complex quantum mechanics
has been known for some time \cite{wootters1986quantum, wootters1990local}.
It concerns multipartite systems: the state of the
multipartite system is completely determined when
the states on the subsystems including the correlations
have been identified. This means that state tomography of
the multipartite system can be performed by simultaneous
measurements at all the subsystems. 
Since this feature has become a candidate for a last postulate
to perform the final step in the quantum-mechanical 
reconstruction process and to rule out the non-complex versions,
it has been named \emph{local tomography}
\cite{barnum2019strongly, Bar_Wil2014locTom, Barnum2016post_cl, barrett2007information,
chiribella2011informational, de2012deriving, Hardy_Foliable2011, masanes2011derivation}.

In the present paper, local tomography is first defined 
in a quantum logical setting and is then postulated only for
bipartite systems consisting of two copies of the same system. 
This postulate is sufficient to prove the need for the complex numbers
in quantum mechanics, when the Jordan algebraic setting is settled.

The main result can be applied to complete 
reconstructions of quantum mechanics 
that first derive the formally real Jordan algebras
\cite{Barnum_2013, barnum2019strongly, e19060253, 
barnum2014higher, nakahira2019derivation, 
nie2020charJordan, wetering2018, van_de_Wetering_2019, 
Wilce_2012, Wilce_2018, Wilce_2019}.
In some of these approaches, local tomography has already been used. 
However, their methods, frameworks and assumptions
differ from those of the present paper and, particularly, 
their local tomography requirement is more restrictive than the quantum logical
definition of local tomography.

Formally real Jordan algebras and probabilistic models of multipartite systems 
that do not necessarily entail local tomography 
are studied in \cite{Barnum_2015, barnum2016composites, wetering2018}.

The formally real Jordan algebras include the so-called spin factors
that involve some peculiarities in the quantum logical setting.
With a spin factor, the maximum number of possible outcomes
in a single measurement is two. Quantum logics with this property
have a very weak mathematical structure. In this case, the geometric methods
of the early quantum logical approaches fail 
\cite{piron1964axiomatique, varadarajan1968and1970}, 
the postulates of \cite{nie2020charJordan} are not satisfied
and the Gleason theorem does not hold 
\cite{bunce1985quantum, gleason1957measures}. 

This paper is based on the theory of Jordan algebras, 
as presented in the monographs \cite{AS02, hanche1984jordan}.
The results needed here are briefly 
sketched in section 4.

The quantum logical structure used in this paper is introduced in the next section.
In section 3, reasonable postulates for a model of a composite system, 
including local tomography,
are defined in the quantum logical setting.
Section 4 provides the brief sketch of the formally real Jordan algebras
and some first simple results that will be needed later.
The final local tomography postulate
in the Jordan algebraic setting, 
a discussion of the spin factors 
and some further results for later use are presented in section 5.
Section~6 prepares some auxiliaries needed to prove the major results
in section~7.
\newpage
\noindent
\large
\textbf{2. Quantum logics}
\\[0,3 cm]
\normalsize
The quantum logic of usual quantum mechanics consists of the observables 
with the simple spectrum $\left\{ 0,1 \right\}$ 
(or, equivalently, the self-adjoint projection operators on the Hilbert space
or the closed linear subspaces of the Hilbert space)
and forms an orthomodular lattice.
Originally, therefore, a quantum logic was mostly assumed to be an 
orthomodular lattice \cite{piron1964axiomatique, varadarajan1968and1970}. 
However, there is no physical motivation 
for the existence of the lattice operations
for propositions that are not compatible, and later a quantum logic 
was often assumed to be an orthomodular partially ordered set only.
This is what we will do here.

In this paper, a \textit{quantum logic} shall be an orthomodular partially ordered set $L$ 
with order relation $\leq$, 
smallest element $0$, largest element $\mathbb{I} \neq 0$ and an orthocomplementation $'$.
This means that the following conditions are satisfied by the $p,q \in L$:
\begin{enumerate}
\item[(a)] $ q \leq p$ \textit{implies} $p' \leq q'$.
\item[(b)] $(p')' = p$.
\item[(c)] $p \leq q'$ \textit{implies that} $p \vee q$, \textit{the supremum of} $p$ \textit{and} $q$, \textit{exists}.
\item[(d)] $p \vee p' = \mathbb{I}$.
\item[(e)] Orthomodular law: $q \leq p$ \textit{implies} $p = q \vee (p \wedge q')$.
Here, $p \wedge q$ denotes the infimum of $p$ and $q$, which exists iff $p' \vee q'$ exists.
\end{enumerate}
The elements of the quantum logic are called \textit{propositions}. 
A proposition $p$ is called \textit{minimal}, 
if there is no proposition $q$ with $q \leq p$ and $0 \neq q \neq p$.
The minimal propositions are also called \textit{atoms} in the common literature.
Two propositions $p$ and $q$ are \textit{orthogonal}, if $p \leq q'$ or, equivalently, $q \leq p'$;
in this case, $p \vee q$ will be noted by $p + q$ in the following.
The interpretation of this mathematical terminology is as
follows: orthogonal events are exclusive, $p'$ is the negation of
$p$, and $p + q := p \vee q $ is the disjunction of the exclusive events $p$ and $q$.

The mathematical structure used in classical probability theory 
is the \emph{Boolean lattice}, and it can be
expected that those subsets of the quantum logic $L$ 
that are Boolean lattices
behave classically. Therefore, 
two propositions $p_1$ and $p_2$ in $L$
shall be called \emph{compatible}, if 
there is a subset of $L$ that is a Boolean lattice
and that includes both propositions $p_1$ and $p_2$.
This is equivalent to the existence of three pairwise orthogonal
propositions $q_1, q_2, q_3 \in L$ such that
$p_1 = q_1 + q_2$ and $ p_2 = q_2 + q_3$.

A \emph{state} allocates probability values to the propositions 
and is a non-negative function $\mu$ on the quantum logic $L$ such that
$\mu(\mathbb{I}) = 1$ and $\mu(p+q) = \mu(p) + \mu(q)$ for any two orthogonal 
propositions $p$ and $q$ in $L$.

Multipartite systems play an important role in quantum theory, 
but a general model of a composite system is not available 
in the quantum logical setting 
\cite{Aerts1982, Aerts1984, Foulis_Randall1981, Ischi2000}. 
In the following section,
reasonable postulates for such a model will be defined; 
a general quantum logical solution satisfying these postulates 
does not exist, but it will be seen in section 7 
that the tensor product used 
in quantum mechanics with the complex Hilbert space 
fulfils the postulates 
and that this distinguishes the complex case from other cases.
\\[0,3 cm]
\large
\textbf{3. Composite systems and local tomography}
\\[0,3 cm]
\normalsize
In quantum mechanics as well as in classical probability theory,
the global state of a multipartite system can be 
determined completely by specifying joint probabilities of outcomes
for measurements performed simultaneously on
each subsystem. However, this is not possible
if a real version of quantum mechanics is considered
instead of the usual one using the complex Hilbert space.
This quantum-mechanical feature has been known 
for a long time \cite{wootters1986quantum, wootters1990local}.
When it later got significance
in different approaches to reconstruct quantum mechanics,
it was called \emph{local tomography} 
\cite{barnum2019strongly, Bar_Wil2014locTom, barrett2007information,
chiribella2011informational, de2012deriving, Hardy_Foliable2011, masanes2011derivation}.
These approaches
often use local tomography as a last postulate to
finally segregate usual quantum theory with the complex Hilbert space
from the real and other versions.
The concept of local tomography shall now be transferred to
the quantum logical setting and the postulates for the
model of a composite system shall be introduced.

When the quantum logics $L_1$ and $L_2$
are used as models for two single systems and
the quantum logic $L_{12}$ is used as the model
for the composite system consisting of these two systems,
there shall be a map $\otimes: L_1 \times L_2 \rightarrow L_{12}$ such that
the following conditions hold:
\itshape
\begin{enumerate}
\item[\emph{(C1)}]
$\mathbb{I} \otimes \mathbb{I} $ is the largest element of $L_{12}$.
\item[\emph{(C2)}] 
For $p_1 \in L_1$ and $p_2 \in L_2$, 
$p_1 \otimes p_2 = 0$ holds in $L_{12}$ iff $p_1 = 0$ or $p_2 = 0$.
\item[\emph{(C3)}] 
If $p_1$ and $q_1$ are orthogonal in $L_1$ or 
if $p_2$ and $q_2$ are orthogonal in $L_2$,
then $p_1 \otimes p_2$ and $q_1 \otimes q_2$ are orthogonal in $L_{12}$.
\item[\emph{(C4)}] 
If $p_1$ and $q_1$ are orthogonal in $L_1$, then 
$ (p_1 +q_1) \otimes p_2 = p_1 \otimes p_2 + q_1 \otimes p_2 $ for any $p_2 \in L_2$.

If $p_2$ and $q_2$ are orthogonal in $L_2$, then 
$ p_1  \otimes (p_2 + q_2) = p_1 \otimes p_2 + p_1 \otimes q_2 $ for any $p_1 \in L_1$.
\item[\emph{(C5)}] 
If $\mu_1$ and $\mu_2$ are states on $L_{12}$ such that
$\mu_1(p_1 \otimes p_2) = \mu_2(p_1 \otimes p_2)$ for all $p_1 \in L_1$, $p_2 \in L_2$,
then $\mu_1 = \mu_2$.
\end{enumerate}
\normalfont
The maps $p_1 \rightarrow p_1 \otimes \mathbb{I}$ and $p_2 \rightarrow \mathbb{I} \otimes p_2$ provide embeddings of $L_1$ and $L_2$ in $L_{12}$. For every $p_1 \in L_1 $ and every $p_2 \in L_2 $, 
$p_1 \otimes \mathbb{I}$ and $\mathbb{I} \otimes p_2$ are compatible in $L_{12}$, since
$p_1 \otimes \mathbb{I} = p_1 \otimes p_2 + p_1 \otimes p_2'$, 
$\mathbb{I} \otimes p_2 = p_1 \otimes p_2 + p_1' \otimes p_2$ and 
$p_1 \otimes p_2$, $p_1' \otimes p_2$, $p_1 \otimes p_2'$ are 
pairwise orthogonal in $L_{12}$. 
\newpage
\noindent
\textbf{Lemma 3.1.} 
\itshape If the first four conditions \emph{(C1--4)} hold and 
$p_k,q_k$ is a compatible pair in $L_k$ for $k = 1,2$, then 
$p_1 \otimes p_2$ and $q_1 \otimes q_2$ are compatible in $L_{12}$.
\normalfont
\\[0,3 cm]
Proof. Suppose that $p_k,q_k$ are compatible in $L_k$ for $k = 1,2$.
Then there are pairwise orthogonal propositions $x_1,x_2,x_3 \in L_1$ and
pairwise orthogonal propositions $y_1,y_2,y_3 \in L_2$ such that
$p_1 = x_1 + x_2$, $q_1 = x_2 + x_3$, $p_2 = y_1 + y_2$, $q_2 = y_2 + y_3$.
Consider 
$z_1 := x_1 \otimes y_1 + x_2 \otimes y_1 + x_1 \otimes y_2$, 
$z_2 := x_2 \otimes y_2$, 
$z_3 := x_2 \otimes y_3 + x_3 \otimes y_2 + x_3 \otimes y_3$.
These are pairwise orthogonal propositions in $L_{12}$ with
$z_1 + z_2 = p_1 \otimes p_2$ and $ z_2 + z_3 = q_1 \otimes q_2$.
\hfill $\square$
\\[0,3 cm]
\textbf{Lemma 3.2.}
\itshape
If the first four conditions \emph{(C1--4)} hold and 
$q_k \leq p_k$ with $p_k , q_k \in L_k$, $k = 1,2$, then 
$q_1 \otimes q_2 \leq p_1 \otimes p_2$ in $L_{12}$.
\\[0,3 cm]
\normalfont
Proof.
Suppose $q_k \leq p_k$ for $k = 1,2$. Then
$p_k = q_k + p_k \wedge q_k'$ and 
$p_1 \otimes p_2 = q_1 \otimes q_2 + q_1 \otimes (p_2 \wedge q_2') 
+ (p_1 \wedge q_1') \otimes q_2 + (p_1 \wedge q_1') \otimes (p_2 \wedge q_2')
\geq q_1 \otimes q_2$.
\hfill $\square$
\\[0,3 cm]
(C1--4) are general, purely algebraic postulates for modeling a composite system 
in the quantum logical setting, and 
(C5) is basically the local tomography postulate.

The only-if part of (C2) [$p_1 \otimes p_2 = 0 \Rightarrow p_1 = 0$ \textit{or} $p_2 = 0$]
means that the embeddings of $L_1$ and $L_2$ in $L_{12}$ are \emph{logically independent}.
Logical independence is usually defined for 
von Neumann subalgebras \cite{hamhalter1997statistical, redei1995logical} 
and becomes a necessary and sufficient condition 
for the \emph{C*-independence} 
of two commuting von Neumann subalgebras \cite{redei1995logical}; 
C*-independence was introduced by Haag and Kastler in the framework of algebraic quantum field theory \cite{haag1964algebraic}.

A model for a composite system that satisfies the postulates (C1--5) 
does not exist generally, but only for an important subclass of the quantum logics, 
which will be identified later.

Instead of only two, three or 
more systems could be combined into a single one. The subsystems could be very different or 
could be copies of the same system. In the following, local tomography will be 
postulated only for a composite system consisting of two copies of the same system; 
this means that a quantum logic $L$ shall satisfy the above postulates with $L_1 = L_2 = L$.
\\[0,3 cm]
\large
\textbf{4. Jordan algebras}
\\[0,3 cm]
\normalsize
A \textit{Jordan algebra} is a linear space $A$ with a commutative bilinear product $\circ$ 
that satisfies the identity $(x^{2} \circ y) \circ x = x^{2} \circ (y \circ x)$ for $x,y \in A$.
A Jordan algebra over the real numbers is called \textit{formally real}, if 
$x^{2}_{1} + ... + x^{2}_{m} = 0$ implies $x_1 = ... = x_m = 0$ for any $x_1,...,x_m \in A$
and any positive integer $m$. 
The formally real Jordan algebras were introduced, 
studied and classified in \cite{von1933algebraic}.
In the finite-dimensional case, 
they coincide with the so-called JB-algebras and JBW-algebras \cite{AS02, hanche1984jordan}
and with the Euclidean Jordan algebras \cite{barnum2016composites, Faraut, Wilce_2018}.

Each finite-dimensional formally real Jordan algebra $A$ possesses 
a multiplicative identity $\mathbb{I}$ and a natural order relation
such that $a^{2} \geq 0$ holds for all $a \in A$;
the system of the idempotent elements 
$$ L_A := \left\{ p \in A : p^{2} = p \right\} $$
becomes a quantum logic with 
$p' := \mathbb{I} - p$. In the following, $L_A$ will be called 
the quantum logic of the Jordan algebra $A$, and the elements of $L_A$
will be called propositions.

For each finite-dimensional formally real Jordan algebra $A$, there are
two characteristic numbers. The first is the 
\emph{dimension} of $A$ as a real vector space, 
which will be denoted by $n_A$ in the following.
The second is the \emph{rank} of $A$; this is the maximal number of pairwise orthogonal minimal propositions in $L_A$,
which will be denoted by $k_A$ in the following. 
Then $$\mathbb{I} = \sum^{k_A}_{j=1} p_j ,$$ with
$k_A$ pairwise orthogonal minimal propositions $p_1,p_2,...,p_{k_A}$.
Moreover, $k_A \leq n_A$; the identity $k_A = n_A$ means
$$A = \sum^{k_A}_{j=1} \mathbb{R} p_j ,$$
and holds iff the Jordan product is associative.

Two elements $x$ and $y$ in a Jordan algebra $A$
are said to \emph{operator-commute}, if 
$x \circ (y \circ z) = y \circ (x \circ z)$ for all $z \in A$.
The \emph{center} of $A$ consists of all elements 
that operator-commute with every element of $A$
and becomes an associative subalgebra of $A$.
$A$ is simple (irreducible) iff its center is $\mathbb{R} \mathbb{I}$.
Moreover, any two operator-commuting elements of $A$
lie in a joint associative subalgebra of $A$.

In the study of Jordan algebras, an important role is played by the co-called
\emph{Jordan triple product}, which is defined as 
$\left\{x,y,z\right\} := x\circ(y\circ z) - y\circ(z\circ x) + z\circ(x\circ y)$ 
for three elements $x,y,z$ in a Jordan algebra $A$.
In the case of the special Jordan product $x \circ y := (xy + yx)/2$
in an associative algebra, the identity 
$\left\{x,y,x\right\} = xyx$ holds.

A Jordan algebra $A$ is said to be the direct sum of the subalgebras
$A_1$ and $A_2$ ($A = A_1 \oplus A_2$) if $A$ is the direct sum 
of $A_1$ and $A_2$ as linear spaces and if $a_1 \circ a_2 = 0$ 
holds for all $a_1 \in A_1$, $a_2 \in A_2$.
Every finite-dimensional formally real Jordan algebra decomposes 
into a direct sum of simple (irreducible) subalgebras; 
these are either one-dimensional (the real numbers), spin factors 
(a spin factor is characterized by $k_A = 2$ and $n_A \geq 3$)
or can be represented 
as algebras of the Hermitian $k \times k$-matrices
over the real numbers $\mathbb{R}$, complex numbers $\mathbb{C}$, 
quaternions $\mathbb{H}$ with $k = 3,4,5,...$ or 
over the octonions $\mathbb{O}$ with $k=3$ only
\cite{hanche1984jordan, von1933algebraic}.
The product of the matrices $x,y$ is given by $x \circ y := (xy + yx)/2$.
These Jordan matrix algebras are denoted by
$ H_k(\mathbb{R}) $, $ H_k(\mathbb{C}) $, $ H_k(\mathbb{H}) $ and $ H_3(\mathbb{O}) $.
Note that the indexes $k$ and $3$ coincide with the rank
$k_A$ of these algebras.

In \cite{nie2020charJordan}, four postulates for a quantum logic $L$
have been presented that are satisfied if and only if $L$ is the 
quantum logic $L_A$ of some finite-dimensional formally real Jordan algebra $A$,
the decomposition of which into simple algebras does not include spin factors.
This means that,
in this case, the mathematical structure of the quantum logic $L_A$
of the algebra $A$ is so rich that the complete algebra $A$ 
(with its linear and multiplicative structure)
can be recovered from $L_A$ (with the order relation $\leq$ 
and the orthocomplementation $'$).

Let $A$ be a finite-dimensional formally real Jordan algebra.
A linear functional $\mu : A \rightarrow \mathbb{R}$ 
is called \emph{positive}, if 
$\mu(a) \geq 0$ holds for all $a \in A$ with $a \geq 0$.
A positive linear functional $\mu$ with $\mu(\mathbb{I})=\mathbb{I}$
is called a \emph{state} on $A$; its restriction to $L_A$ 
becomes a state on this quantum logic.
By Gleason's theorem \cite{gleason1957measures} 
and its extension to Jordan algebras \cite{bunce1985quantum}, 
each state $\mu$ on the quantum logic $L_A$
has a unique linear extension to $A$,
if the decomposition of $A$ into simple subalgebras does not include spin factors.
In this case, the states on the quantum logic $L_A$
can be identified with the states on $A$.
When spin factors are included, however, it is necessary
to distinguish between the linear states on the algebra $A$
and the states on quantum logic $L_A$.
Each linear functional $\mu$ on $A$ can be written as
$\mu = t_1 \mu_1 - t_2 \mu_2$ with non-negative 
real number $t_1,t_2$ and states $\mu_1, \mu_2$ on $A$.

The next lemma is a collection of facts from the theory of the formally real Jordan algebras
\cite{AS02, hanche1984jordan, von1933algebraic}
that will be needed in the following sections. 
\\[0,3 cm]
\textbf{Lemma 4.1.} 
\itshape
Let $A$ be a finite-dimensional formally real Jordan algebra and 
let $p$ and $q$ be propositions in its quantum logic $L_A$.
\begin{enumerate}
\item[\emph{(i)}]
$ 0 \leq \left\{p,q,p\right\} \leq p$, and
$ 0 \leq \left\{p,a,p\right\}$ for any $ a \in A $ with $ a \geq 0 $.
\item[\emph{(ii)}]
$p$ is minimal iff $\left\{p,A,p\right\} = \mathbb{R} p$ and $p \neq 0$.
\item[\emph{(iii)}]
$p$ and $q$ are orthogonal iff $p \circ q = 0$ 
iff $\left\{p,q,p\right\} = 0$ iff $\left\{q,p,q\right\} = 0$.
Moreover, $p \leq q$ iff $p \circ q = p$ iff $\left\{p,q,p\right\} = p$ 
iff $\left\{q,p,q\right\} = p$.
\item[\emph{(iv)}]
Orthogonal propositions in $L_A$ operator-commute in $A$.
\item[\emph{(v)}] \emph{Spectral theorem:}
Every $a \in A$ can be written as $ a = \sum_{j} r_j q_j$ with pairwise
orthogonal minimal proposition $q_j \in L_A$ and $r_j \in \mathbb{R}$, $j = 1,2,..,m$.

$\left\{p,A,p\right\}$ is a Jordan subalgebra of $A$ and its elements $a$ 
can be written as $ a = \sum_{j} r_j q_j$ with pairwise
orthogonal minimal proposition $q_j \leq p$ and $r_j \in \mathbb{R}$, $j = 1,2,..,m$.
\item[\emph{(vi)}]
If $\mu$ is a linear state on $A$ with $\mu(p) = 1$, then 
$\mu(\left\{p,a,p\right\}) = \mu(a)$ for all $a \in A$.
\end{enumerate}
\normalfont
\textbf{Definition 4.2.} 
\itshape
Let $p$ be a minimal proposition in the quantum logic $L_A$ 
of a finite-dimensional formally real Jordan algebra $A$.
For each $a \in A$ there is $r_a \in \mathbb{R}$ with 
$ \left\{ p,a,p \right\} = r_a p$ by lemma \emph{4.1 (ii)}.
Thus $p$ induces a linear state 
$\mathbb{P}_p : A \rightarrow \mathbb{R}$, $ a \rightarrow r_a$
with $\mathbb{P}_p (p) = 1$.
\\[0,3 cm]
\normalfont
The positivity of $\mathbb{P}_p$ follows from lemma 4.1(i).
For a minimal proposition $p$ and any other proposition $q$ in the quantum logic $L_A$ 
of a finite-dimensional formally real Jordan algebra $A$,
lemma 4.1(iii) immediately yields:
\begin{enumerate}
\item[] 
$\mathbb{P}_p (q) = 0$ iff $p$ and $q$ are orthogonal, and
\item[] 
$\mathbb{P}_p (q) = 1$ iff $p \leq q$.
\end{enumerate}
\textbf{Lemma 4.3.}
\itshape
Let $A$ be a finite-dimensional formally real Jordan algebra and $a \in A$ with
$\mathbb{P}_p (a) = 0$ for all minimal minimal propositions $p \in L_A$. Then $a = 0$.
\\[0,3 cm]
Proof.
\normalfont
By lemma 4.1(v), $ a = \sum_{j} r_j q_j$ with pairwise
orthogonal minimal proposition $q_j \in L_A$, $r_j \in \mathbb{R}$, $j = 1,2,..,m$,
and $0 = \mathbb{P}_{q_j} (a) = r_j$ for each $j$ implies $a = 0$.
\hfill $\square$
\\[0,3 cm]
\textbf{Lemma 4.4.}
\itshape
Let $A$ be a finite-dimensional formally real Jordan algebra and 
let $p$ and $q$ be propositions in its quantum logic $L_A$.
Then $p$ and $q$ operator-commute in the Jordan algebra $A$ iff 
$p$ and $q$ are compatible in the quantum logic $L_A$.
\\[0,3 cm]
\normalfont
Proof.
First suppose that $p$ and $q$ operator-commute in $A$,
which means that both lie in a joint associative subalgebra of $A$.
Then $ p \circ q = (p \circ q)^{2} $ and $ x_1 := p \circ q $,
$x_2 := p - x_1$, $x_3 := q - x_1$
are pairwise orthogonal propositions in $L_A$.
This means that $p$ and $q$ are compatible in $L_A$.

Now suppose that $p$ and $q$ are compatible in $L_A$.
This means that there are pairwise orthogonal 
propositions $x_1, x_2, x_3 \in L_A$ with
$p = x_1 + x_2$ and $q = x_2 + x_3$.
By lemma 4.1(iv), $x_1, x_2, x_3$ pairwise operator-commute
and, therefore, $p$ and $q$ operator-commute in $A$.
\hfill $\square$
\\[0,3 cm]
Two orthogonal propositions $p,q$ in the quantum logic $L_A$ 
of a Jordan algebra $A$ are said to be \emph{strongly connected}, 
if there is an element $x \in \left\{p,A,q\right\}$ such that $x^{2} = p + q$.
If $A$ is formally real, finite-dimensional and simple, each pair of
orthogonal minimal propositions is strongly connected \cite{hanche1984jordan}.
\\[0,3 cm]
\textbf{Lemma 4.5.}
\itshape
If $p_1$ and $p_2$ are any minimal propositions in the quantum logic $L_A$
of a simple finite-dimensional formally real Jordan algebra $A$, then
there is a further minimal proposition $q \in A$ 
that is orthogonal to neither $p_1$ nor $p_2$.
\\[0,3 cm]
\normalfont
Proof. If $p_1$ and $p_2$ are not orthogonal, choose $q := p_1$ or $q := p_2$.
Now assume that $p_1$ and $p_2$ are orthogonal. Since $A$ is simple,  $p_1$ and $p_2$ are 
strongly connected and hence $\left\{p_1,A,p_2\right\} \neq \left\{ 0 \right\}$.
$A$ is the linear span of its minimal propositions and, therefore, 
$\left\{p_1,q,p_2\right\} \neq 0 $ must hold for at least one minimal proposition $q$.
Then $0 \neq p_1 \circ (q \circ p_2) - q \circ (p_2 \circ p_1) + p_2 \circ (p_1 \circ q) 
= p_1 \circ (q \circ p_2) + p_2 \circ (p_1 \circ q)$. 
Since $p_1$ and $p_2$ operator-commute, the last two summands are identical; therefore
each one cannot vanish and thus $ p_1 \circ q \neq 0 \neq p_2 \circ q $. This means
that $q$ is orthogonal to neither $p_1$ nor $p_2$.
\hfill $\square$
\newpage
\noindent
\large
\textbf{5. The local tomography postulate}
\\[0,3 cm]
\normalsize
Generally, a physical system is not monolithic, 
but comprises many smaller subsystems. 
In the simplest multipartite case, it consists of two copies 
of the same subsystem, and it will turn out 
that the consideration of this case is sufficient 
to rule out the non-complex versions of quantum mechanics.

The local tomography postulate for the quantum logic $L_A$ of a 
finite-di\-men\-sion\-al formally real Jordan algebra $A$
can now be presented:
\itshape
\begin{enumerate}
\item[\emph{(LT)}] There shall be 
another finite-dimensional formally real Jordan algebra $A^{2}$ 
and a map $\otimes: L_A \times L_A \rightarrow L_{A^{2}}$
such that the five conditions \emph{(C1--5)} in section \emph{3} hold
with $L_1 = L_2 = L_A$ and $L_{12} = L_{A^{2}}$.
\end{enumerate}
\normalfont
This quantum logical version of local tomography is 
not as restrictive as the prevalent version in other papers, where 
the map $\otimes$ is immediately postulated to exist 
as a bilinear map on $A \times A$ or as a biconvex map on
$\left[0,\mathbb{I}\right] \times \left[0,\mathbb{I}\right]$
and not only as a biadditive map on $L_A \times L_A$. The convex set
$\left[0,\mathbb{I}\right] = \left\{a \in A : 0 \leq a \leq \mathbb{I}\right\}$
is the so-called \emph{effect space}, playing a major role 
in some approaches to reconstruct quantum mechanics.
With (LT), the linear extension of the map $\otimes$ to $A \times A$ 
is not immediately given and will be derived later in lemma 5.1; 
in proposition 7.1, it will
be shown that it is one-to-one.

If $A = H_k(\mathbb{C})$, then $H_{k^{2}}(\mathbb{C})$ 
is the tensor product $A \otimes A$ and, in this case, the
quantum logic $L_A$ satisfies (LT).
However, such a solution is not available in the real case, since
$H_k(\mathbb{R}) \otimes H_k(\mathbb{R})$ and $ H_{k^{2}}(\mathbb{R})$ 
have different dimensions [$(k(k+1)/2)^{2} \neq k^{2}(k^{2}+1)/2 $ for $k > 1$].

Moreover, the case $H_2(\mathbb{C})$ is special;
(LT) is satisfied not only with the usual map
$\otimes: L_{H_2(\mathbb{C})} \times L_{H_2(\mathbb{C})} \rightarrow L_{H_4(\mathbb{C})}$,
but also with many other maps, which
can be generated by the discontinuous automorphisms of this quantum logic.
For every $q \in L_{H_2(\mathbb{C})}$ with $0 \neq q \neq \mathbb{I}$,
an example of such a discontinuous automorphism 
can be defined in the following way:
$\pi_q (p) := p$ for $p \in L_{H_2(\mathbb{C})}$ with $q \neq p \neq q'$ 
and $\pi_q (q) := q'$, $\pi_q (') := q$.
An alternative map $\otimes_{\pi}$ is then given by
$p_1 \otimes_{\pi} p_2 := \pi(p_1) \otimes \pi(p_2) \in L_{H_4(\mathbb{C})}$
for $p_1,p_2 \in L_{H_2(\mathbb{C})}$.

Furthermore, the quantum logics $L_{A_1}$ and $L_{A_2}$ 
of two spin factors $A_1$ and $A_2$ are isomorphic 
iff $L_{A_1}$ and $L_{A_2}$ have the same cardinality: 
the minimal propositions in $L_{A_1}$ can be organized 
into the sets $\left\{p,p'\right\}$ and so can those in $L_{A_2}$.
The only rules that an isomorphism $\pi$ from $L_{A_1}$ to $L_{A_2}$
has to obey are that $\pi(0) = 0$, $\pi(\mathbb{I}) = \mathbb{I}$ and
that the two-element subsets of $L_{A_1}$ are mapped 
one-to-one to those of $L_{A_2}$. This only requires 
that $L_{A_1}$ and $L_{A_2}$ have the same cardinality.

The quantum logic $L_A$ of every finite-dimensional spin factor $A$
has the cardinality of the continuum, becomes thus
isomorphic to $L_{H_2(\mathbb{C})}$ and satisfies (LT) with 
$A^{2} = H_4(\mathbb{C})$.
The origin of this unattractive feature lies in the weak algebraic structure 
of the quantum logic of a spin factor, since it contains only minimal propositions
and $0$ and $\mathbb{I}$. Therefore, spin factors will often play
an exceptional role in the following.
\newpage
\noindent
\textbf{Lemma 5.1.} 
\itshape
Assume that \emph{(LT)} holds for the quantum logic $L_A$ of a 
finite-dimensional formally real Jordan algebra $A$.
\begin{enumerate}
\item[\emph{(i)}]
$A^{2}$ is the linear hull of $\left\{ p \otimes q : p,q \in L_A \right\}$.
\item[\emph{(ii)}]
Assume furthermore that the decomposition of $A$ into simple subalgebras does
not include spin factors. 

\emph{(a)} Suppose $ p_0 = \sum_{i=1}^{k} r_i p_i$ 
and $ q_0 = \sum_{j=1}^{l} s_j q_j$ in $A$
with $p_0,p_1,p_2,...,p_k,$ $q_0,q_1,q_2,...,q_l \in L_A$ and 
$ r_i, s_j \in \mathbb{R}$. Then 
$$ p_0 \otimes q_0 = \sum_{i,j=1}^{k,l} r_i s_j \  p_i \otimes q_j $$ 
in $A^{2}$. 

\emph{(b)} The map $ \otimes : L_A \times L_A \rightarrow A^{2} $ can be extended
in a unique way 
to a bilinear map $ \otimes : A \times A \rightarrow A^{2} $.
\end{enumerate}
\normalfont
Proof.
(i) Assume that $\left\{ p \otimes q : p,q \in L_A \right\}$
does not generate $A^{2}$. Then there is a linear functional
$\mu : A^{2} \rightarrow \mathbb{R}$ with $\mu \neq 0$ and 
$\mu(p \otimes q) = 0$ for all $p,q \in L_A$.
It can be written as $\mu = t_1 \mu_1 - t_2 \mu_2$ with non-negative 
real numbers $t_1,t_2$ and two states $\mu_1, \mu_2$ on $A^{2}$.
Since $0 = \mu(\mathbb{I} \otimes \mathbb{I}) = t_1 - t_2$, we get $t_1 = t_2$.
Either $t_1 = t_2 = 0$ and $\mu = 0$ or
$ \mu_1 (p \otimes q) = \mu_2 (p \otimes q) $ for all $p,q \in L_A$.
In the second case, (CS5) implies $\mu_1 = \mu_2$ and thus $\mu = 0$
in both cases, which yields the desired contradiction.

(ii) (a) 
Let $\mu: A^{2} \rightarrow \mathbb{R}$ be a linear functional on $A^{2}$; 
it can again be written as $\mu = t_1 \mu_1 - t_2 \mu_2$ with non-negative 
real number $t_1,t_2$ and states $\mu_1, \mu_2$ on $A^{2}$.

Define 
$\mu^{p_0}_{1}$ on $L_A$ by $\mu^{p_0}_{1} (q) := \mu_1 (p_0 \otimes q)$ for $q \in L_A$.
By the extension of Gleason's theorem to Jordan algebras \cite{bunce1985quantum}, 
$\mu^{p_0}_{1}$ has a unique linear extension to $A$. Therefore
$$\mu^{p_0}_{1}(q_0) = \sum_{j=1}^{l} s_j \mu^{p_0}_1 (q_j).$$
For each $j$ now define 
$\mu^{q_j}_{1}$ on $L_A$ by $\mu^{q_j}_{1} (p) := \mu_{1}(p \otimes q_j)$ for $p \in L_A$.
Again each $\mu^{q_j}_{1}$ has a unique linear extension to $A$ and 
$$\mu^{q_j}_{1} (p_0) = \sum_{i=1}^{k} r_i \mu^{q_j}_{1}(p_i).$$
Then
\begin{align*}
\mu_{1}(p_0 \otimes q_0) 
&= \mu^{p_0}_{1}(q_0) 
= \sum_{j=1}^{l} s_j \mu^{p_0}_{1}(q_j) 
= \sum_{j=1}^{l} s_j \mu_1 (p_0 \otimes q_j) \\
&= \sum_{j=1}^{l} s_j \mu^{q_j}_{1}(p_0) 
= \sum_{j=1}^{l} s_j \sum_{i=1}^{k} r_i \mu^{q_j}_{1}(p_i) \\
&= \sum_{i,j=1}^{k,l} r_i s_j \mu_1(p_i \otimes q_j).
\end{align*}
The identity
$$\mu_{2}(p_0 \otimes q_0) = \sum_{i,j=1}^{k,l} r_i s_j \mu_2(p_i \otimes q_j)$$
follows for $\mu_2$ in the same way, and we get for $\mu = t_1 \mu_1 - t_2 \mu_2$
$$ \mu(p_0 \otimes q_0) = \sum_{i,j=1}^{k,l} r_i s_j \mu(p_i \otimes q_j) 
= \mu \left(\sum_{i,j=1}^{k,l} r_i s_j \  p_i \otimes q_j \right).$$
Since this holds for any linear functional $\mu$ on $A^{2}$,
we get the desired identity of lemma 5.1(ii)(a).

(ii) (b) 
$A$ is the linear hull of $L_A$. 
Pick a basis in $A$ with elements $p_i \in L_A$, $i = 1,...,n_A$,
and map the pair $\sum_{i=1}^{n_A} r_i p_i $ and $ \sum_{j=1}^{n_A} s_j p_j $ in $A$ 
to $ \sum_{i,j=1}^{n_A} r_i s_j \ p_i \otimes p_j$ in $A^{2}$.
By (ii)(a), the pairs $(p_0,q_0)$ with $p_0,q_0 \in L_A$
are mapped to $p_0 \otimes q_0$ in $A^{2}$.
The uniqueness of the extension is then implied by (i).
\hfill $\square$
\\[0,3 cm]
\textbf{Lemma 5.2.} 
\itshape
Assume that \emph{(LT)} holds for the quantum logic $L_A$ of a 
finite-dimensional formally real Jordan algebra $A$.
If the propositions $p_1,p_2 \in L_A$ lie in the center of $A$, then
$p_1 \otimes p_2$ lies in the center of $A^{2}$.
\\[0,3 cm]
\normalfont
Proof. Suppose that $p_1,p_2 \in L_A$ lie in the center of $A$.
By lemma 4.4, $p_1$ and $p_2$ are compatible with every
proposition in $L_A$ and, by lemma 3.1, $p_1 \otimes p_2$ is
compatible with every $q_1 \otimes q_2$ in $L_{A^{2}}$, $q_1,q_2 \in L_A$.
By lemma 4.4, $q_1 \otimes q_2$ operator-commutes with every such $q_1 \otimes q_2$
and therefore with every element of $A^{2}$ by lemma 5.1(i).
\hfill $\square$
\\[0,3 cm]
\textbf{Lemma 5.3.} 
\itshape
If \emph{(LT)} holds for the quantum logic $L_A$ of a 
finite-dimensional formally real Jordan algebra $A$ and if
$A$ is the direct sum of the subalgebras $A_1$ and $A_2$,
then \emph{(LT)} holds for each quantum logic $L_{A_1}$ and $L_{A_2}$.
\\[0,3 cm]
\normalfont
Proof. Let $\mathbb{I}_1$ and $\mathbb{I}_2$ be the 
multiplicative identities of $A_1$ and $A_2$. 
They lie in the center of $A$ and 
$\mathbb{I}_1 + \mathbb{I}_2 = \mathbb{I}$. 
By lemma 5.2, $\mathbb{I}_i \otimes \mathbb{I}_j $
lies in the center of $A^{2}$ for $i,j = 1,2$.
The sum of these four propositions is 
$\mathbb{I} \otimes \mathbb{I}$ and therefore
$$ A^{2} = \oplus_{i,j = 1,2} \left\{\mathbb{I}_i \otimes \mathbb{I}_j, A^{2}, \mathbb{I}_i \otimes \mathbb{I}_j\right\}.$$
For $p,q \in L_{A_1}$, we have
$p,q \in L_{A} $ and $p,q \leq \mathbb{I}_1$.
By lemma 3.2, $p \otimes q \leq \mathbb{I}_1 \otimes \mathbb{I}_1$ in~$A^{2}$
and therefore 
$ p \otimes q \in \left\{\mathbb{I}_1 \otimes \mathbb{I}_1, A^{2}, \mathbb{I}_1 \otimes \mathbb{I}_1\right\}$
by lemma 4.1(iii).
The restriction of the map $\otimes$
to $L_{A_1} \times L_{A_1}$ then satisfies (CS1--4);
$L_1 = L_2 = L_{A_1}$ and $L_{12}$ is the quantum logic of the formally real Jordan algebra 
$\left\{\mathbb{I}_1 \otimes \mathbb{I}_1, A^{2}, \mathbb{I}_1 \otimes \mathbb{I}_1\right\}$.

Now let $\mu_1$ and $\mu_2$ be states on this quantum logic 
with $\mu_1 (p_1 \otimes q_1)$ = $\mu_2 (p_1 \otimes q_1)$ for all $p_1,q_1 \in L_{A_1}$.
Define states $\nu_1$ and $\nu_2$ on $L_{A^{2}}$ by
$$\nu_k(x) := \mu_k(\left\{ \mathbb{I}_1 \otimes \mathbb{I}_1, x, \mathbb{I}_1 \otimes \mathbb{I}_1 \right\})$$
for $x \in L_{A^{2}}$, $k=1,2$. Note that
$\left\{ \mathbb{I}_1 \otimes \mathbb{I}_1, x, \mathbb{I}_1 \otimes \mathbb{I}_1 \right\}$
is idempotent for $x \in L_{A^{2}}$ 
because $\mathbb{I}_1 \otimes \mathbb{I}_1$ lies in the center of $A^{2}$.

Suppose $p,q \in L_{A}$. Then 
$p = p_1 + p_2$ and $q = q_1 + q_2$ with $p_1,q_1 \in L_{A_1}$, $p_2,q_2 \in L_{A_2}$
and hence $p \otimes q= \sum_{i,j} p_i \otimes q_j$.
By (C3) and lemma 4.1(iii),
$\left\{\mathbb{I}_1 \otimes \mathbb{I}_1, p_i \otimes q_j,\mathbb{I}_1 \otimes \mathbb{I}_1 \right\} = 0$ for $i \neq 1$ or $j \neq 1$.
Thus 
$\left\{\mathbb{I}_1 \otimes \mathbb{I}_1, p \otimes q, \mathbb{I}_1 \otimes \mathbb{I}_1 \right\} 
= \left\{ \mathbb{I}_1 \otimes \mathbb{I}_1, p_1 \otimes q_1 ,\mathbb{I}_1 \otimes \mathbb{I}_1 \right\}
= p_1 \otimes q_1 $ by lemma 3.2. Therefore,
$$\nu_1(p \otimes q) 
= \mu_1(p_1 \otimes q_1)
= \mu_2(p_1 \otimes q_1)
= \nu_2(p \otimes q).$$
Since this holds for all $p,q \in L_{A}$ and
$L_A$ satisfies (LT), we get $\nu_1 = \nu_2$ 
on $L_{A^{2}}$. By lemma 4.1(iii), the restriction of $\nu_k$ to the quantum logic of 
$\left\{\mathbb{I}_1 \otimes \mathbb{I}_1, A^{2}, \mathbb{I}_1 \otimes \mathbb{I}_1\right\}$
coincides with $\mu_k$, $k = 1,2$, and we get
$\mu_1 = \mu_2$.
That $L_{A_2}$ satisfies (LT) follows in the same way.
\hfill $\square$
\\[0,3 cm]
\large
\textbf{6. Auxiliaries}
\\[0,3 cm]
\normalsize
In this section, it shall always be assumed
that $A$ is a finite-dimensional formally real Jordan algebra,
that its decomposition into simple subalgebras does
not include spin factors
and that its quantum logic $L_A$ satisfies (LT). 
\\[0,3 cm]
\textbf{Lemma 6.1.} 
\itshape
Suppose 
$ \left\{ p_1,q_1,p_1 \right\} = r_1 p_1$ and $ \left\{ p_2,q_2,p_2 \right\} = r_2 p_2$ 
with $p_1, p_2,$ $q_1, q_2 \in L_A$ and $r_1, r_2 \in \mathbb{R}$. Then 
$$ \left\{ p_1 \otimes p_2 , q_1 \otimes q_2 , p_1 \otimes p_2 \right\} = r_1 r_2 \ p_1 \otimes p_2 $$
in $A^{2}$. 
\\[0,3 cm]
\normalfont
Proof. 
If $r_1 = 0$, then $q_1$ and $p_1$ are orthogonal by lemma 4.1(iii); 
therefore, $p_1 \otimes p_2$ and 
$q_1 \otimes q_2$ are orthogonal by (C3) and both sides of the identity above 
become $0$. 

Now assume $r_1 > 0$, and
let $\mu$ be a state on $ A^{2}$ with $ \mu(p_1 \otimes p_2) = 1$.
Then 
$ 1 \geq \mu ( \mathbb{I} \otimes p_2 ) = \mu ( p_1 \otimes p_2 ) + \mu ( {p_1}' \otimes p_2 ) \geq 1 $
and thus $ \mu ( \mathbb{I} \otimes p_2 ) = 1$. Therefore 
\begin{center}
$ \mu_1(a) := \mu ( a \otimes p_2 )$, $a \in A $,
\end{center}
defines a state on $A$ with $ \mu_1(p_1) = 1$;
note that $a \otimes p_2$ is defined by lemma 5.1(ii)(b).
By lemma 4.1 (vi), $$\mu_1(q_1) = \mu_1(\left\{ p_1,q_1,p_1 \right\}) = r_1 \mu_1(p_1) = r_1 .$$
Now define, again using lemma 5.1(ii)(b), 
\begin{center}
$\mu_2 (b) := \frac{1}{r_1} \mu(q_1 \otimes b)$, $b \in A$.
\end{center}
Then $\mu_2 (p_2) = \mu_1(q_1)/r_1 = 1 $. 
Furthermore, $p_1 \otimes p_2$ and $q_1 \otimes {p_2}'$
are orthogonal by (C3) and 
$1 = \mu(\mathbb{I} \otimes \mathbb{I} )$
$\geq \mu (p_1 \otimes p_2 + q_1 \otimes {p_2}')$
$= \mu (p_1 \otimes p_2) + \mu(q_1 \otimes {p_2}')$
$= 1 + \mu(q_1 \otimes {p_2}') \geq 1$
implies $\mu(q_1 \otimes {p_2}') = 0$,
hence $\mu_2 (p_2') = 0$
and therefore $\mu_2 (\mathbb{I}) = \mu_2 (p_2) + \mu_2 (p_2') = 1$.
This means that $\mu_2$ becomes a state on $A$ with $\mu_2(p_2)=1$.
Once more applying lemma 4.1(vi) yields
$$ \mu_2(q_2) = \mu_2(\left\{p_2,q_2,p_2\right\}) = r_2 \mu_2(p_2) = r_2 .$$ 
Therefore
$$\mu(q_1 \otimes q_2) = r_1 \mu_2(q_2) = r_1 r_2 .$$
Now let $\nu$ be any state on $ A^{2} $ with $\nu(p_1 \otimes p_2) > 0$. Then
$$\mu(x) := \frac{1}{\nu(p_1 \otimes p_2)} \nu(\left\{p_1 \otimes p_2,x,p_1 \otimes p_2\right\}), \ x \in A^{2}, $$
becomes a state with $\mu(p_1 \otimes p_2) = 1 $ and, as shown before,
$$\mu(q_1 \otimes q_2) = r_1 r_2 .$$
This means
$$\nu(\left\{p_1 \otimes p_2,q_1 \otimes q_2,p_1 \otimes p_2\right\}) = r_1 r_2 \ \nu(p_1 \otimes p_2).$$
This identity holds for all states $\nu$ including those with $\nu(p_1 \otimes p_2) = 0$, 
since then $0 \leq \nu(\left\{p_1 \otimes p_2,q_1 \otimes q_2,p_1 \otimes p_2\right\}) \leq \nu(p_1 \otimes p_2) = 0$ by lemma 4.1(i).
Therefore 
$$ \left\{ p_1 \otimes p_2 , q_1 \otimes q_2 , p_1 \otimes p_2 \right\} = r_1 r_2 \ p_1 \otimes p_2 .$$
\hfill $\square$
\\[0,3 cm]
\textbf{Lemma 6.2.}
\itshape
Let $p_1, p_2$ be minimal propositions in $L_A$.
\begin{enumerate}
\item[\emph{(i)}]
$p_1 \otimes p_2$ is a minimal proposition in $L_{A^{2}}$.
\item[\emph{(ii)}]
$\mathbb{P}_{p_1 \otimes p_2} ({q_1 \otimes q_2}) = \mathbb{P}_{p_1} (q_1) \mathbb{P}_{p_2} (q_2)$
for all $q_1, q_2 \in L_A$.
\item[\emph{(iii)}]
If $q_1, q_2$ are propositions in $L_A$ such that
$p_1 \otimes p_2$ and $q_1 \otimes q_2$ are orthogonal in $L_{A^{2}}$, 
then either $p_1$ and $q_1$ or $p_2$ and $q_2$ or both pairs are orthogonal in $L_A$.
\end{enumerate}
\normalfont
Proof. First note that $p_1 \otimes p_2 \neq 0$ by (C2) in section 3.

(i) Let $q_1,q_2 \in L_A$.
By lemma 4.1(ii), 
$ \left\{ p_1,q_1,p_1 \right\} = r_1 p_1$ and $ \left\{ p_2,q_2,p_2 \right\} = r_2 p_2$
with reals $r_1, r_2$.
By lemma 6.1,
$ \left\{ p_1 \otimes p_2 , q_1 \otimes q_2 , p_1 \otimes p_2 \right\} = r_1 r_2 \ p_1 \otimes p_2 $. 

By lemma 5.1(i), 
$A^{2}$ is the linear hull of 
$\left\{ q_1 \otimes q_2 \ : \ q_1, q_2 \in L_A \right\}$,
and we get 
$ \left\{ p_1 \otimes p_2 , A^{2} , p_1 \otimes p_2 \right\} = \mathbb{R} \  p_1 \otimes p_2$.
By lemma 4.1(ii), $p_1 \otimes p_2$ becomes minimal in $L_{A^{2}}$.

(ii) This follows from Definition 4.2, lemma 6.1 and lemma 6.2(i).

(iii) Suppose that 
$p_1 \otimes p_2$ and $q_1 \otimes q_2$ are orthogonal in $A^{2}$.
By lemma 6.2(ii) then
$\mathbb{P}_{p_1} (q_1) \mathbb{P}_{p_2} (q_2) = \mathbb{P}_{p_1 \otimes p_2} ({q_1 \otimes q_2}) = 0$.
Therefore $\mathbb{P}_{p_1} (q_1) = 0$ and $p_1$ and $q_1$ are orthogonal,
or $\mathbb{P}_{p_2} (q_2) = 0$ and $p_2$ and $q_2$ are orthogonal.
\hfill $\square$
\\[0,3 cm]
\textbf{Lemma 6.3.} 
\itshape
Suppose that $A$ is simple, and let $p_1, p_2, q_1, q_2 \in L_A$ 
be minimal propositions. Then
$p_1 \otimes p_2$ and $q_1 \otimes q_2$ belong to the same 
simple direct summand of $A^{2}$.
\\[0,3 cm]
\normalfont
Proof. By lemma 4.5, there are minimal propositions $e_1$ and $e_2$ in $L_A$ 
such that $e_1$ is orthogonal to neither $p_1$ nor $q_1$ and
$e_2$ is orthogonal to neither $p_2$ nor $q_2$.
By lemma 6.2(iii), $e_1\otimes e_2$ is orthogonal 
to neither $p_1\otimes p_2$ nor $q_1\otimes q_2$ in $A^{2}$.
Therefore, both $p_1\otimes p_2$ and $q_1\otimes q_2$ belong to 
the same direct summand as $e_1\otimes e_2$, because minimal
propositions from different direct summands are orthogonal.
\hfill $\square$
\\[0,3 cm]
\textbf{Lemma 6.4.} 
\itshape
If $A$ is simple, then $A^{2}$ is simple too.
\normalfont
\\[0,3 cm]
Proof. Let $p_i$, $i = 1,2,...,k_A$, be pairwise orthogonal 
minimal propositions in $L_A$ with 
$\sum p_i = \mathbb{I}$.
By lemma 6.3, the $p_i \otimes p_j$, $i,j = 1,2,...,k_A$,
all belong to the same simple direct summand of $A^{2}$.
This direct summand then includes 
$$\sum^{k_A}_{i,j=1} p_i \otimes p_j 
= \left(\sum^{k_A}_{i=1} p_i\right) \otimes \left(\sum^{k_A}_{j=1} p_j\right) 
= \mathbb{I} \otimes \mathbb{I} ,$$
and thus becomes the complete algebra $A^{2}$. Therefore,
$A^{2}$ is simple.
\hfill $\square$
\\[0,3 cm]
Findings, analogous to the last two lemmas above,
are contained in \cite{wetering2018},
but another framework and other assumptions are used;
local tomography is not postulated and 
the real version of quantum mechanics remains included.
\\[0,3 cm]
\large
\textbf{7. Results}
\\[0,3 cm]
\normalsize
\textbf{Proposition 7.1.} 
\itshape
Let $A$ be a finite-dimensional formally real Jordan algebra
such that its decomposition into simple subalgebras does
not include spin factors and its quantum logic $L_A$ satisfies \emph{(LT)}.
The two characteristic numbers, dimension and rank, 
then factorize in the following way:
$n_{A^2} = {n_A}^2$ and $k_{A^2} = {k_A}^2$.
\\[0,3 cm]
\normalfont
Proof. 
If $\mathbb{I} = \sum^{k_A}_{i=1} p_i$ in $A$
with pairwise orthogonal minimal propositions $p_i$ in $L_A$, then 
$\mathbb{I} \otimes \mathbb{I} = \sum^{k_A}_{i,j=1} p_i \otimes p_j$ in $A^2$. 
By lemma 6.2(i), the $ p_i \otimes p_j$ are minimal propositions in $L_{A^{2}}$
and therefore $k_{A^2} = {k_A}^2$.

Now let $q_i$, $i = 1,2...,n_A$, be a basis of $A$ such that each $q_i$
is a proposition in $L_A$. It shall be shown that the
$q_i \otimes q_j$, $i,j = 1,2...,n_A$, become a basis in $A^2$.

Assume $0 = \sum_{ij} r_{ij} (q_i \otimes q_j)$ with $ r_{ij} \in \mathbb{R}$.
For any two minimal propositions $p_1,p_2 \in L_A$ then by lemma 6.2(i) and (ii)
\begin{align*}
0 &= \mathbb{P}_{p_1 \otimes p_2} \left( \sum_{ij} r_{ij} (q_i \otimes q_j) \right)
= \sum_{ij} r_{ij} \mathbb{P}_{p_1 \otimes p_2} ( q_i \otimes q_j ) \\
&= \sum_{ij} r_{ij} \mathbb{P}_{p_1} ( q_i ) \mathbb{P}_{p_2} ( q_j )
= \mathbb{P}_{p_2} \left( \sum_{ij} r_{ij} \mathbb{P}_{p_1} ( q_i ) \ q_j \right).
\end{align*}
Since this holds for all minimal propositions $p_2 \in L_A$, we get 
by lemma 4.3
$$ 0 = \sum_{ij} r_{ij} \mathbb{P}_{p_1} ( q_i ) q_j.$$
The linear independence of the $q_j$ implies
$$ 0 = \sum_i r_{ij} \mathbb{P}_{p_1} ( q_i ) = \mathbb{P}_{p_1} \left( \sum_i r_{ij} q_i \right) ,$$
for each $j$.
Since this still holds for all minimal propositions $p_1 \in L_A$, we get by lemma 4.3 again
$$ 0 = \sum_i r_{ij} q_i ,$$
for each $j$, and the linear independence of the $q_i$ implies 
$ r_{ij} = 0 $ for all $i,j = 1,2...,n_A$.

Thus we have the linear independence of the 
$q_i \otimes q_j$, $i,j = 1,2...,n_A$, in $A^2$.
By lemma 5.1, these elements become a basis of $A^2$ and therefore $n_{A^2} = {n_A}^2$.

\hfill $\square$
\\[0,3 cm]
\textbf{Proposition 7.2.} 
\itshape
The quantum logics of $H_{k}(\mathbb{R})$, $H_{k}(\mathbb{H})$ with $k \geq 3$ and $H_{3}(\mathbb{O})$ 
do not satisfy \emph{(LT)}. 
\\[0,3 cm]
\normalfont
Proof.
Note that the dimensions of 
$H_{k}(\mathbb{R}), \ H_{k}(\mathbb{C}), \ H_{k}(\mathbb{H})$ and $H_{3}(\mathbb{O})$ 
are $k(k+1)/2, \ k^{2}, \ k(2 k - 1)$ and $27$, respectively.
Let $A$ be one of the algebras 
$H_{k}(\mathbb{R})$, $H_{k}(\mathbb{H})$ with $k \geq 3$ or $H_{3}(\mathbb{O})$.

If (LT) were satisfied, $A^{2}$ would be a simple formally real Jordan algebra 
and thus a matrix algebra with rank
$k_{A^{2}} = k^{2}$ and, for $A = H_{3}(\mathbb{O})$, $k_{A^{2}} = 9$.
This follows from lemma 6.4 and proposition 7.1. For the dimension $n_{A^{2}}$
of $A^{2}$, proposition 7.1 implies the following:
\begin{enumerate}
\item[]
$n_{A^{2}} = k^{2}(k+1)^{2}/4$ if $A = H_{k}(\mathbb{R})$,
\item[]
$n_{A^{2}} = k^{2}(2 k - 1)^{2}$ if $A = H_{k}(\mathbb{H})$, and
\item[]
$n_{A^{2}} = 27^{2} = 729$ if $A = H_{3}(\mathbb{O})$.
\end{enumerate}
The dimension $n_{A^{2}}$ must equal
either $k^{2}(k^{2}+1)/2$ if $A^{2}$ were a real matrix algebra,
or $k^{4}$ if $A^{2}$ were a complex matrix algebra,
or $k^{2}(2 k^{2} - 1)$ if $A^{2}$ were a matrix algebra over the quaternions.
Simple calculations show that each
case is impossible for $k \geq 3$.
With $A = H_{3}(\mathbb{O})$, $n_{A^{2}} = 729$ 
would have to be identical to one of the dimensions of
$H_{9}(\mathbb{R})$, $H_{9}(\mathbb{C})$, $H_{9}(\mathbb{H})$, 
but these are 45, 81 and 153, respectively.
Therefore, (LT) cannot hold.
\hfill $\square$
\\[0,3 cm]
\textbf{Theorem 7.3.}
\itshape
Let $A$ be a finite-dimensional formally real Jordan algebra
such that its decomposition into simple subalgebras does
not include spin factors.
Its quantum logic $L_A$ satisfies \emph{(LT)} iff $A$ is the 
the self-adjoint part of a C*-algebra.
\\[0,3 cm]
\normalfont
Proof.
$A$ is the direct sum of algebras of the following types: $H_k (K)$ with 
$K = \mathbb{R}, \mathbb{C}, \mathbb{H}$ and $k \geq 3$, $H_3 (\mathbb{O})$ 
and $\mathbb{R}$.
If (LT) holds, lemma 5.3 and proposition 7.2 rule out all the cases 
with $K = \mathbb{R},\mathbb{H}$ and $H_3(\mathbb{O})$, 
and the direct sum can include only 
the cases $H_k(\mathbb{C})$ and $\mathbb{R}$. This means that $A$ is 
the self-adjoint part of a C*-algebra.

Vice versa, if $A$ is the self-adjoint part of a C*-algebra, 
let $A^{2}$ be the self-adjoint part
of the tensor product of two copies of this 
C*-algebra; (LT) is then fulfilled
with the usual embedding $L_A \times L_A \ni (p_1,p_2) \rightarrow p_1 \otimes p_2$.
\hfill $\square$
\\[0,3 cm]
\textbf{Theorem 7.4.}
\itshape
The quantum logic $L_A$ of a finite-dimensional formally real Jordan algebra $A$
satisfies \emph{(LT)} iff $A$ is the direct sum of
the self-adjoint part of a C*-algebra and spin factors.
\\[0,3 cm]
\normalfont
Proof. Suppose that (LT) holds. By lemma 5.3 and proposition 7.2,
only the complex matrix algebras, spin factors and $\mathbb{R}$ 
can occur in the decomposition of $A$ into simple subalgebras
and $A$ becomes the direct sum of the self-adjoint part of a C*-algebra 
and spin factors.

Now suppose that $A$ is the direct sum of
the self-adjoint part of a C*-algebra and spin factors.
The quantum logic of each spin factor is isomorphic 
to the quantum logic of $H_2(\mathbb{C})$. This is 
the self-adjoint part of the C*-algebra 
consisting of the $2 \times 2$-matrices.
Using these isomorphisms, $A^{2}$ can again be constructed
as the self-adjoint part of the tensor product of C*-algebras.
\hfill $\square$
\\[0,3 cm]
In \cite{hardy2001quantum, van_de_Wetering_2019, zaopo2012information}, 
the identity $n_{A^2} = {n_A}^2$ in proposition 7.1
for the dimensions of $A^2$ and $A$
is not derived, but simply introduced as one of the postulates.
Results analogous to those of this section and, particularly, a proof 
close to the one of proposition 7.2 are included in 
\cite{van_de_Wetering_2019}.
However, the theory of the sequential product spaces 
and the postulates used in \cite{van_de_Wetering_2019}
differ a lot from the quantum logical approach. This is why 
the role of the spin factors becomes different in the two approaches
and all except the complex $2 \times 2$-matrices 
can be ruled out in \cite{van_de_Wetering_2019}.
\\[0,3 cm]
\large
\textbf{8. Conclusion}
\\[0,3 cm]
\normalsize
Theorem 7.3 gets particularly interesting when it is combined
with the results of \cite{nie2020charJordan}. 
Four postulates for a quantum logic $L$ were presented there 
that are satisfied 
iff $L$ is the quantum logic of a 
finite-dimensional formally real Jordan algebra,
the decomposition of which into simple subalgebras does
not include spin factors. 
Local tomography becomes the perfect add-on to 
these four postulates in order to finally bring us to 
common quantum mechanics, when (LT) is replaced by the following version:
\begin{enumerate}
\item[(LT')]
\emph{For the quantum logic $L$, there shall be 
another quantum logic $L^{2}$ 
and a map $\otimes: L \times L \rightarrow L^{2}$
such that the five conditions \emph{(C1--5)} in section \emph{3} hold
with $L_1 = L_2 = L$ and $L_{12} = L^{2}$, and both $L$ and $L^{2}$
shall satisfy the four postulates of} \cite{nie2020charJordan}. 
\end{enumerate}
We then get the following result:
\newpage
\begin{enumerate}
\item[]
\emph{A quantum logic $L$ satisfies the four postulates of 
\emph{\cite{nie2020charJordan}} and \emph{(LT')}}
\begin{center}
\emph{if and only if}
\end{center}
\emph{$L$ is the quantum logic formed by the self-adjoint projections 
in a finite-dimensional C*-algebra that does not include 
the complex $2 \times 2$-matrix algebra as a direct summand.}
\end{enumerate}
The postulate (LT) can also
be applied to complete other
reconstructions of quantum mechanics 
that first derive the formally real Jordan algebras
\cite{Barnum_2013, barnum2019strongly, e19060253, 
barnum2014higher, nakahira2019derivation, 
nie2020charJordan, wetering2018, van_de_Wetering_2019, 
Wilce_2012, Wilce_2018, Wilce_2019}.
Some of these reconstructions already use local tomography,
but a stronger version (see section 5) 
which could be replaced by the less restrictive (LT). Moreover,
their methods, frameworks and assumptions differ
from those of the present paper, and they often do not include 
the non-simple (reducible) algebras.

Erroneously, in \cite{niestegge2004composite}, a postulate 
that is not even fulfilled by quantum mechanics 
with the complex Hilbert space was considered 
for the model of a composite system \cite{nie-erratum2010}.
It requires local tomography and something more.
The present paper demonstrates that pure local tomography 
and an additional requirement for the map $\otimes$
are the right replacement for it in order to derive the need for the 
complex numbers in quantum mechanics.
The new requirement for the map $\otimes$
that was not used in \cite{niestegge2004composite}
is the only-if part of (C2) in section 3 (the logical independence).

In this paper, only finite-dimensional formally real 
Jordan algebras have been considered. Their infinite-dimensional analogues
are the so-called JBW-algebras \cite{AS02, hanche1984jordan}. 
The idempotent elements of 
a JBW-algebra form a quantum logic and the 
local tomography postulate can easily be transferred. 
An interesting question now becomes whether (LT) 
distinguishes the self-adjoint parts of the von Neumann algebras 
among the JBW-algebras in the same way as 
in the finite-dimensional case (theorems 7.3 and 7.4).
This problem cannot be tackled with the mathematical tools
used here, since the structure theory and classification
of the JBW-algebras with infinite dimension 
are a lot richer than the finite-dimensional case.

In order to derive the quantum-mechanical need for the C*-algebras
or von Neumann algebras,
an alternative to local tomography, which does not involve 
bipartite or multipartite systems, is the notion of 
\emph{dynamical correspondence}. It concerns the mathematical
model of the dynamical evolution of a single system and is motivated by
the quantum-mechanical feature that the dynamical group is generated by the
Hamilton operator \cite{AS02, barnum2014higher, nie2015dyn}.
Further mathematical alternatives to distinguish the self-adjoint parts 
of the C*-algebras or von Neumann algebras
among the JB-algebras or JBW-algebras
(3-\emph{ball property} and \emph{orientability} \cite{AS02, AS01})
lack a physical or probabilistic motivation.

Theorem 7.4 includes the quantum logics of spin factors,
but involves a high degree of ambiguity 
in this case, to which the two-dimensional Hilbert space 
and the single qubit belong. The ambiguity cannot be
circumvented in the quantum logical setting 
because of the discontinuous automorphisms and isomorphisms
of the quantum logics of spin factors (see section 5).
This may be regarded as a drawback of the quantum logical approach.
Its benefits lie in its fundamentality. Instead of presuming
linear or convex structures from the very beginning, 
it starts with the more basic structure of the
quantum logic, which is a generalization of the classical Boolean lattice. 
Nonetheless, in the case of the quantum logic 
of a formally really Jordan algebra without spin factors as simple subalgebras,
this structure is rich enough to recover from it the complete algebra 
with the linear and multiplicative structure \cite{nie2020charJordan}.

The spin factors are unproblematic
in the case of other approaches such as the grand work by Alfsen and Shultz,
the contemporary operational probabilistic theories or the sequential product spaces,
since they put convex structures (the state space or the effect space)
and their characteristics into the focus.
\\[0,3 cm]
\emph{Data accessibility.} All of the data are contained within the paper.
\newline
\emph{Competing interests.} I declare I have no competing interest.
\newline
\emph{Funding.} I received no funding for this study.

\bibliographystyle{abbrv}
\bibliography{Literatur}
\end{document}